\documentclass[%
 reprint,%
 aip,apl,%
]{revtex4-1}

\usepackage{bm}%
\usepackage[colorlinks=true,linkcolor=blue]{hyperref}%
\expandafter\ifx\csname package@font\endcsname\relax\else
 \expandafter\expandafter
 \expandafter\usepackage
 \expandafter\expandafter
 \expandafter{\csname package@font\endcsname}%
\fi
\usepackage{graphicx}
\usepackage{dcolumn}

\begin{document}

\preprint{APS/123-QED}

\title{Spin-to-Charge Conversion in 2D Electron Gas and Single-layer Graphene Devices}

\author{J. G. G. S. Ramos}
\affiliation{Departamento de F\'{\i}sica, Universidade Federal da Para\'iba, 58297-000 Jo\~ao Pessoa, Para\'iba, Brazil}

\author{T. C. Vasconcelos}
\affiliation{Departamento de F\'{\i}sica, Universidade Federal da Para\'iba, 58297-000 Jo\~ao Pessoa, Para\'iba, Brazil}

\author{A. L. R. Barbosa}
\email{anderson.barbosa@ufrpe.br}
\affiliation{Departamento de F\'{\i}sica, Universidade Federal Rural de Pernambuco, 52171-900, Recife, PE, Brazil }

\date{\today}

\begin{abstract}
We investigate the spin-to-charge conversion emerging from a mesoscopic device connected to multiple terminals. We obtain analytical expressions to the characteristic coefficient of spin-to-charge conversion which are applied in two kinds of ballistic chaotic quantum dots at low temperature. We perform analytical diagrammatic calculations in the universal regime for two-dimensional electron gas and single-layer graphene with strong spin-orbit interaction in the universal regime. Furthermore, our analytical results are confirmed by numerical simulations. Finally, we connect our analytical finds to recent experimental measures giving a conceptual explanation about the apparent discrepancies between them.
\end{abstract}


\pacs{Valid PACS appear here}
\maketitle

 
\section{Introduction}
Charge-to-spin and spin-to-charge conversions are key aspects of spintronics and have been the focus of a large number of experimental and theoretical investigations ~\cite{Niimil,Sinova,Rojas-Sanchez,Shen,Bhattacharjee1}. They are experimentally performed in materials with large spin-orbit interaction (SOI) as some normal metals ~\cite{Sinova,YZhang,Huang1}, two-dimensional electron gas (2DEG) ~\cite{Rojas-Sanchez,Shen,Zhang,Isasa,Stano,Nichele}, single-layer graphene (SLG) ~\cite{Ohshima,Mendes,Dushenko,Torres} and topological insulators (TIs) ~\cite{Rojas-Sanchez1}. In bulk materials, the forme case is known as spin Hall effect (SHE), for which a longitudinal charge current crossing a region with SOI gives rise to a transversal pure spin current. The latter case is known as inverse spin Hall effect (ISHE), which is the Onsanger reciprocal of the SHE ~\cite{jaquodbuttiker}, i.e., a longitudinal pure spin current produces a transversal charge current. The efficiency of the ISHE is commonly identified with a characteristic coefficient, the spin Hall angle $\Theta_{\mathrm{SHE}}$ is given by the ratio between the charge current ($\mathrm{A/m}$) and the spin current ($\mathrm{A}/\mathrm{m}^2$) densities, $\Theta_{\mathrm{SHE}}t/2=j^c/j^s$. It is also assumed the spin diffusion length typically much larger than the material thickness ($t$).

Differently of bulk materials, the charge-to-spin conversion mechanism in 2DEG is determined by the Rashba-Edelstein effect (REE) ~\cite{Rashba,Edelstein,Aronov}. The Hamiltonian that describes the SOI in the REE can be written as $\mathrm{H}_{\mathrm{R}} = \alpha_{\mathrm{R}} (\mathbf{k}\times \hat{z})\cdot\mathbf{\sigma}$, where $\alpha_{\mathrm{R}}$ is the Rashba coefficient whereas $ \hat{z}$ and $\mathbf{\sigma}$ are, respectively, the unit vector perpendicular to the plane of the 2DEG and the Pauli matrices vector. Similarly to the SHE/ISHE reciprocity, there is a reciprocal effect of the REE which is known as inverse Rashba-Edelstein effect (IREE) ~\cite{Rojas-Sanchez,Shen}. The efficiency of the IREE can be also obtained from the ratio between charge current and spin current densities, generating the IREE characteristic conversion parameter,  $\lambda_{\mathrm{IREE}}=j^c/j^s$. However, it is proportional to the momentum relaxation time $\tau$ and the Rashba coefficient as $\lambda_{\mathrm{IREE}}=\alpha_{\mathrm{R}}\tau/\hbar$.

The first experimental measure of $\lambda_{\mathrm{IREE}}$ was reported in Ref.~\cite{Rojas-Sanchez}. Rojas-Sanchez {\it et.al.} injected a pure spin current using a spin pumping from NiFe layer into a 2DEG surface formed on Ag/Bi interface, and obtained values of $\lambda_{\mathrm{IREE}}$ ranging from $0.2$ to $0.33\; \mathrm{nm}$. However, Zhang  {\it et.al.}  ~\cite{Zhang}, after some time, performed the same experiment in samples of NiFe/Ag/Bi and obtained values of $\lambda_{\mathrm{IREE}}$ in the interval $0.11\pm 0.02 \; \mathrm{nm}$, which are two or three times smaller compared with the results obtained by Ref.~\cite{Rojas-Sanchez}. Therefore, the relevant question emerges: what are the physical concepts behind the experimental discrepancies between Refs.~\cite{Rojas-Sanchez} and ~\cite{Zhang}?

Furthermore, Mendes {\it et.al.} ~\cite{Mendes} reported an experiment that inject a pure spin current using a spin pumping from a ferromagnetic insulator composed of yttrium iron garnet into the SLG. They interpreted the spin-to-charge conversion in SLG as a manifestation of the IREE and obtained that the value of characteristic conversion parameter is $\lambda_{\mathrm{IREE}}\approx 10^{-3} \; \mathrm{nm}$, that is two orders of magnitude smaller than was obtained by Ref.~\cite{Rojas-Sanchez}. Nevertheless, Ohshima {\it et. al.} ~\cite{Ohshima}, which in an earlier moment performed an experiment with similar results of Ref.~\cite{Mendes}, interpreted spin-to-charge conversion in SLG as ISHE. The latter interpretation is also supported by Dushenko {\it et. al.} ~\cite{Dushenko}. Despite this disagreement, it is  correct to assert that $j^c/j^s \approx 10^{-3} \; \mathrm{nm}$ for SLG from Ref.~\cite{Mendes}. Again arise an essential question: what is the physical concepts behind the experimental discrepancies between Refs.~\cite{Rojas-Sanchez} and ~\cite{Mendes}?
 
In this work, we investigate the spin-to-charge conversion using the Landauer-B\"uttiker formulation \cite{Buttiker}. We obtain expressions to the characteristic coefficient of spin-to-charge conversion in this framework. Moreover, we apply the analytical results to the universal ballistic chaotic quantum dots with SOI at low temperature.  We calculate the ensemble average and sample-to-sample universal fluctuations of the characteristic transport coefficient using the random matrix theory (RMT) framework ~\cite{Mehta,Verbaarschot}, i.e., admitting a spin-orbit interaction time much smaller than the dwell time of charges in the quantum dot, $\tau_{\mathrm{dwell}}\gg\tau_{\mathrm{SOI}}$ ~\cite{Jacquod}. Our analytical diagrammatic results are obtained to very distinct mesoscopic devices, those whose 2DEG electrons are wave functions described by the Schr\"odinger equation and also to relativistic Dirac wave functions (SLG).  Furthermore, our results are connected with the experimental measures of Refs.~\cite{Rojas-Sanchez,Zhang, Mendes} giving a conceptual explanation about the apparent discrepancies between the empirical results in the scattering framework. Moreover, our analytical results are confirmed by numerical simulations found on Mahaux-Weidenm\"uller model of a ballistic chaotic quantum dot  ~\cite{Weidenmuller}.

\section{Theoretical Scattering Framework } 

\begin{figure}
\centering
\includegraphics[width=8.0cm,,height=4.0cm]{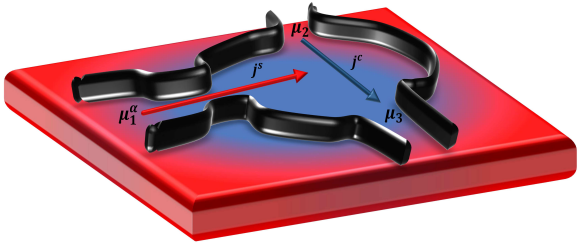}
\caption{A ballistic chaotic quantum dot connected to three electron reservoirs that are set with electrochemical potential and/or spin accumulation given by $\mu_i^0=eV_i$ and  $\mu_i^\beta=(\mu_i^\uparrow-\mu_i^\downarrow)/2$, respectively. The spin-to-charge conversion take place if the reservoir $1$ injects a pure spin current $j^s$ inside the device, which given rise a pure charge current $j^c$ between the leads $2$ and $3$.}
\label{Imagem1}
\end{figure}

We consider a 2D mesoscopic device connected to three electronic reservoirs by ideal leads. Inside the device, the electrons flow under the influence of a large SOI at low temperature. A schematic design of the device is depicted in the Fig.(\ref{Imagem1}). The reservoir indexed by 1 is hold by a spin pump that inject a pure spin current in the 2D device. Therefore, the ISHE or the IREE generates a pure charge current through the leads indexed by 2 and 3. 

We consider the Landauer-B\"uttiker formulation \cite{Buttiker} to describe the  spin-to-charge current conversion through the 2D device.  The spin and charge currents in the $i$th lead are defined as $I_i^\alpha= I_i^{\uparrow}- I_i^{\downarrow}$ and  $I_i^0 =I_i^{\uparrow} + I_i^{\downarrow}$, respectively, with $I_i^{\uparrow}$ ($I_i^{\downarrow}$) is the charge current with spin-up (down) crossing the $i$th lead ~\cite{Jacquod}. The general expression of current is given by Ref. ~\cite{jaquodbuttiker} as following
\begin{equation}
I_{i}^\alpha=\frac{e}{h}\sum_{\beta}\left(2N_i\delta_{\alpha\beta}-\tau_{ii}^{\alpha\beta}\right)\mu_i^{\beta}-\frac{e}{h}\sum_{j\neq{i},\beta}\tau^{\alpha\beta}_{ij}\mu_j^{\beta}.\label{LB}
\end{equation}
The Latin characters indicate the current flow lead index ($i=1,2,3$) whereas the Greek indices were introduced to differentiate charge current ($\alpha=0$) and spin current ($\alpha=x,y,z$). The sum on $j$ and $\beta$ runs over $1,2,3$ and $0,x,y,z$, respectively.  The dimensionless integer $N_i$ is the number of propagating waves channels in the $i$th lead, which is defined as the reason between the lead width ($W_i$) and Fermi wavelength ($\lambda_F$), $N_i \propto W_i/\lambda_F$. 
The electrochemical potential and spin accumulation of $i$th reservoir conected to the $i$th lead are defined as $\mu_i^0=eV_i$ and $\mu_i^\beta=(\mu_i^\uparrow-\mu_i^\downarrow)/2$, respectively, with $\mu_i^\uparrow$ ($\mu_i^\downarrow$) denoting the chemical potential of spin-up (down) along the $\beta$ axis ~\cite{jaquodbuttiker}.  The spin-dependent transmission coefficient $\tau_{ij}^{\alpha\beta}$ gives the spin or charge current transmitted through the $j$th lead from the polarized reservoir $\mu_j^\beta$ to the $i$th lead with corresponding polarized reservoir $\mu_i^\alpha$. It can be obtained from  transmissions and reflections blocks of the corresponding scattering matrix as ~\cite{jaquodbuttiker}
\begin{eqnarray}
\tau_{ij}^{\alpha\beta} &=&\textbf{Tr}\left[\left(\mathcal{S}_{ij}\right)^{\dagger}\sigma^\alpha\mathcal{S}_{ij}\sigma^\beta\right],
\label{tauij}
\end{eqnarray}
with the scattering $\mathcal{S}$-matrix written as
\begin{eqnarray}
\mathcal{S}=
\left[\begin{array}{ccc}
r_{11}&t_{12}&t_{13} \\
t_{21}&  r_{22}&t_{23}\\
t_{31}&  t_{32}&r_{33}
\end{array}\right]. \label{Smatrix}
\end{eqnarray}
Moreover, the $\mathcal{S}$-matrix is unitary ($\mathcal{S}\mathcal{S}^ {\dagger}=1$) and has $2  N_T \times 2 N_T$ dimensional,  with $N_T$ denoting the total number of propagating waves channels given by $N_T = N_1+N_2+N_3$. The $\sigma^0$ and $\sigma^\alpha$ are the identity and Pauli matrices, respectively. The trace of Eq.(\ref{tauij}) is taken over both number of channels and spin spaces.

 Following the Ref.~\cite{Adagideli}, a requirement to the existence of spin current flux through the lead 1 is the set of no charge current flow through it for a specific corresponding electrochemical potential. Therefore, the electrochemical potential of reservoir 1 is set as $\mu_1^0=eV_1$ keeping the charge current null $I_1^0=0$ and a non null spin current $I_1^\alpha\neq 0$ in the Eq.(\ref{LB}). Accordingly, a pure charge current will emerge through the leads $2$ and $3$ as a function of the spin accumulations ($\mu_1^\alpha$) in the reservoir 1, which means, due to conservation law, that $I_2^0=-I_3^0$ in Eq. (\ref{LB}). Hence, this findings imply that all the spin current, the electrochemical potential and the spin accumulations in reservoirs 2 and 3 are analytically set as nulls, $I_2^\alpha =I_3^{\alpha} =0$ and  $\mu_2^0=\mu_3^0=\mu_2^\alpha=\mu_3^\alpha=0$. Using these constraints in the Eq. (\ref{LB}),  we can write the pure spin current as
\begin{equation}
I_{1}^\alpha=\frac{e}{h}\sum_{\beta\neq{0}} \left[\left(2N_1\delta_{\alpha\beta}-\tau_{11}^{\alpha\beta}\right)-\frac{\tau_{11}^{\alpha0}\tau_{11}^{0\beta}}{2N_1-\tau_{11}^{00}}\right]\mu_1^{\beta},\label{I1}
\end{equation}
for $\alpha=x,y,z$, and the pure charge current as
\begin{equation}
I_{i}^0=-\frac{e}{h}\sum_{\beta\neq{0}}\left[\frac{\tau_{i1}^{00}\tau_{11}^{0\beta}}{2N_1-\tau_{11}^{00}}+\tau_{i1}^{0\beta}\right]\mu_1^{\beta},\quad i=2,3.\label{I23}
\end{equation}
 The spin and charge currents, Eqs.(\ref{I1}) and (\ref{I23}), are functions of spin accumulation $\mu_1^\beta$ of reservoir $1$ without dependence on electrochemical potential $\mu_1^0 = eV_1$. Therefore, the spin accumulation tends to zero as the spin and charge currents vanish. Furthermore, we can inspect that the Eq.(\ref{I23}) satisfies the conservation law $\sum_{i=1}^3 I_{i}^0=0$ as expected.  

The Eqs.(\ref{I1}) and (\ref{I23}) allow one to obtain the dimensionless characteristic coefficient $\lambda_\alpha=I^0_i/I_1^\alpha$ of the spin-to-charge conversion as a function of $\tau_{i1}^{\alpha\beta}$ and $\mu_1^\beta$. However, there are two experimental regimes that deserve attention. The first one is taken when only one of the three spin accumulations components of reservoir 1 contributes to spin current or, more specifically, when $\mu_1^\beta=\delta_{\alpha\beta}\mu_1^\alpha$ in Eqs.(\ref{I1}) and (\ref{I23}), for which we obtain
\begin{equation}
\lambda_{\alpha}=\frac{\tau_{i1}^{00}\tau_{11}^{0\alpha}+(2N_1-\tau_{11}^{00})
\tau_{i1}^{0\alpha}}{\tau_{11}^{\alpha0}
\tau_{11}^{0\alpha}-(2N_1-
\tau_{11}^{00})(2N_1-
\tau_{11}^{\alpha\alpha})},\quad i=2,3.\label{C1}
\end{equation}
The second case is taken when the three spin accumulations components of reservoir 1 are symmetric $\mu_1^x=\mu_1^y=\mu_1^z$. Hence, we obtain that
\begin{equation}
\lambda_{\alpha}=\frac{\sum_{\beta}\tau_{i1}^{00}{
\tau_{11}^{0\beta}} + (2N_1-\tau_{11}^{00})
{\tau_{i1}^{0\alpha}}}{
\sum_{\beta}\tau_{11}^{\alpha0}\tau_{11}^{0\beta}+(2N_1-\tau_{11}^{00})
\left[\sum_{\beta\neq\alpha}\tau_{11}^{\alpha\beta}-(2N_1-\tau_{11}^{\alpha\alpha})
\right]}.\label{C2}
\end{equation}
 The Eqs.(\ref{C1}) and (\ref{C2}) can be applied to any type of electronic device. In the next section, we will use these results to investigate two kinds of chaotic quantum dots and to connect our analytical diagrammatic finds with experimental measures of Refs.~\cite{Rojas-Sanchez,Zhang, Mendes}. 

\section{Applied in Chaotic Quantum dot}

\subsection{Chaotic Schr\"odinger Quantum Dot} 

A typical 2D ballistic mesoscopic device, which is schematically depicted in the Fig.(\ref{Imagem1}), is known as ballistic chaotic quantum dot. The wave functions of the electrons through the device are described by the Schr\"odinger equation in such a way that it is better called as ballistic chaotic Schr\"odinger quantum dot (SQD). 

Let us focus on the experimentally relevant case of a coherent SQD, which preserve the time-reversal symmetry (absence of the magnetic field). If there is a strong SOI on the SQD, the scattering matrix of the Eq.(\ref{Smatrix}) is a member of the circular symplectic ensemble in the framework of RMT ~\cite{Mehta}. This allows one to use the standard diagrammatic method ~\cite{Brouwer,RamosBarbosa} to obtain the average and the universal fluctuations of the characteristic coefficients expressed on the Eqs.(\ref{C1}) and (\ref{C2}). 

Firstly, the large $N_T$ analytical calculation of $\lambda_\alpha$ through the diagrammatic method ~\cite{Jacquod,RamosBarbosa2,Jacquod1} renders for both experimental regimes, Eqs.(\ref{C1},\ref{C2}),
\begin{eqnarray}
\left\langle\lambda_{\alpha}\right\rangle &=& \frac{\left\langle I_{i}^0\right\rangle}{\left\langle I_1^\alpha\right\rangle} = 0, \quad \alpha=x,y,z.\label{media}
\end{eqnarray}
 The Eq.(\ref{media}) shows, from the statistic point of view, that when a spin current is pumping inside the chaotic quantum dot there is the same probability to appear a charge current flux from lead 2 to 3 as from lead 3 to 2, and, consequently, its average is null (see the Appendix for further calculation details). However, for a specific mesoscopic device connected to a spin pump reservoir 1, $\lambda_\alpha$ is in general finite, an information encoded on the variance (universal fluctuation) of the characteristic coefficient that we calculate henceforth. Analyzing the Eq.(\ref{C1}) on the first regime, the diagrammatic method render for large $N_T$
\begin{equation}
\mathrm{var}[\lambda_{\alpha}]=\frac{1}{4}\frac{N_2N_3}{N_1(N_2+N_3)N^2_T}. \label{VarLambda}
\end{equation}
The Eq.(\ref{VarLambda}) is our first main result.  It shows that, although the average of characteristic coefficient is null, Eq.(\ref{media}), a single experimental measure of $\lambda_\alpha$ can assume a large amplitude in the universal regime owing the universal current fluctuations intrinsic of all disorder mesoscopic devices. 

The Eq.(\ref{VarLambda}) can be connected with the experimental measure $\lambda_{\mathrm{IERR}}=0.11\pm 0.02 \; \mathrm{nm}$  of Ref.~\cite{Zhang} taking the fully symmetric configuration, $N_i=N$. Therefore, the characteristic coefficient hold a universal fluctuation given by 
\begin{equation}
\mathrm{rms}[\lambda_{\alpha}]= \mathrm{rms}\left[\frac{I^0_i}{I_1^\alpha}\right] =\sqrt{\frac{1}{72} \times \frac{1}{N^2}}\approx 0.11 \times \frac{1}{N} .\label{rmsl}
\end{equation}
 Notice firstly the number of propagating wave channels $N$ is proportional to the reason between the lead width ($W$) and Fermi wavelength ($\lambda_F$), $N\propto W/\lambda_F$. Then, with a very good approximation, we can set henceforth the observable amplitudes in unities of $\lambda_F$ (nm). Furthermore, the charge current per unit width and the spin current per unit width squared can be written, respectively, as $j^c\propto I^0_i/W$ and $j^s\propto I^\alpha_1/W^2$. Accordingly, the universal fluctuation of $\lambda_{\mathrm{IERR}}=j^c/j^s$ in unities of $\lambda_F$ (nm) is $\mathrm{rms}[\lambda_{\mathrm{IERR}}]=\mathrm{rms}[j^c/j^s] \approx \mathrm{rms}[I^0_i/I_1^\alpha] \times W\approx 0.11 \times \lambda_F$, from Eq.(\ref{rmsl}). Hence, we can estimate that $\mathrm{rms}[\lambda_{\mathrm{IERR}}]\approx 0.11 \;\mathrm{nm}$, which is in agreement with Ref.~\cite{Zhang}. 

The diagrammatic calculation to the second regime, Eq.(\ref{C2}), in the limit of large $N_T$ renders
\begin{equation}
\mathrm{var}[\lambda_{\alpha}]=\frac{3}{4}\frac{N_2N_3}{N_1(N_2+N_3)N^2_T}. \label{VarLambda2}
\end{equation}
The Eq.(\ref{VarLambda2}) is our second main result and, fortuitously, three times the one expressed in Eq.(\ref{VarLambda}). Taking the fully symmetric configuration, $N_i=N$, the characteristic coefficient Eq.(\ref{VarLambda2}) supplies the following universal fluctuation 
\begin{equation}
\mathrm{rms}[\lambda_{\alpha}]= \mathrm{rms}\left[\frac{I^0_i}{I_1^\alpha}\right] =\sqrt{\frac{3}{72} \times \frac{1}{N^2}}\approx 0.20 \times \frac{1}{N} .
\end{equation}
Using the same previously dimension analyses of $\lambda_{\mathrm{IERR}}=j^c/j^s$, we can estimate that $\mathrm{rms}[\lambda_{\mathrm{IERR}}]\approx 0.20 \;\mathrm{nm}$.  Furthermore, the total characteristic coefficient can be defined as a linear combination $\lambda =(\lambda_x,\lambda_y,\lambda_z)$, implying that $\mathrm{rms}[\lambda]= \sqrt{\sum_\alpha \mathrm{var}[\lambda_\alpha]}=\sqrt{3\times 3/72 \times 1/N^2}\approx 3 \times 0.11 \times 1/N \approx  0.33 \times 1/N$. Hence, we can estimate that $\mathrm{rms}[\lambda_{\mathrm{IERR}}]\approx 0.33 \;\mathrm{nm}$. Our results are agreement with experimental measure of Ref.~\cite{Rojas-Sanchez}, which predicts values of $\lambda_{\mathrm{IERR}}$ ranging from $0.2 \;\mathrm{nm}$ to $0.33 \;\mathrm{nm}$. The results explaining the difference between experimental measure of Refs.~\cite{Rojas-Sanchez} and  ~\cite{Zhang}.

\subsection{ Chaotic Dirac Quantum Dot} 

The 2D ballistic mesoscopic devices performed with SLG or/and TIs have the electronic wave functions described by the massless Dirac equation of the corresponding relativistic  quantum  mechanics,  instead  of  the  Schr\"odinger equation. Accordingly, an appropriate nomenclature for them is ballistic chaotic Dirac quantum dot (DQD).

As in the previous calculation, the main interest is on the experimentally relevant configuration of a coherent DQD with the time-reversal symmetry (absence magnetic field). Furthermore, the DQD preserve also the sublattice/mirror/chiral symmetry in the Dirac point ~\cite{jaquodbuttiker,Verbaarschot,Hagymasi}. In the presence of large SOI inside the DQD, the scattering matrix of Eq.(\ref{Smatrix}) is a member of the chiral circular symplectic ensemble in the framework of RMT ~\cite{Verbaarschot}, allowing to use the extension of diagrammatic method for Dirac devices ~\cite{Barros,Vasconcelos}.

Again, we expand the ensemble averages of $\lambda_\alpha$ for large $N_T$ using the extension of the diagrammatic method in both the experimental regimes,  (\ref{C1}) and (\ref{C2}). We obtain that $\left\langle\lambda_{\alpha}\right\rangle =0$ as for the SQD. 
However, as previously discussed, the main interest is on the characteristic coefficient variance. Taking the limit of large $N_T$ to the first regime, Eq.(\ref{C1}), the result is 
\begin{equation}
\mathrm{var}[\lambda_{\alpha}]=\frac{1}{8}\frac{N_2N_3}{N_1(N_2+N_3)N^2_T}. \label{VarLambda3}
\end{equation}
 The Eq.(\ref{VarLambda3}) is our third main result and half times the one expressed on the Eq.(\ref{VarLambda}). The factor 2 appears because the chiral symmetric makes the Dirac Hamiltonian two times larger than Schr\"odinger ones. To the second regime, Eq. (\ref{C2}), we obtain
\begin{equation}
\mathrm{var}[\lambda_{\alpha}]=\frac{3}{8}\frac{N_2N_3}{N_1(N_2+N_3)N^2_T}, \label{VarLambda4}
\end{equation}
which is three times the result expressed in the Eq.(\ref{VarLambda3}). 

The Eqs.(\ref{VarLambda3}) and (\ref{VarLambda4}) can establish a relevant connection with the experimental result $\lambda_{\mathrm{IERR}}\approx 10^{-3} \; \mathrm{nm}$  of the Ref.~\cite{Mendes}. 
 Notice firstly, according to Refs.  ~\cite{Gnutzmann,Richter,RamosHusseinBarbosa,Vasconcelos,Guo2008} the chiral universal class is only relevant at low energy, which means in our analytical diagrammatic calculation that the number of propagating wave channels needs to be kept small. If the Fermi energy is set way from zero the Wigner-Dyson e Chiral universal classes lead to the same results.
Hence, the symmetric configuration ($N_i=N$) used in the SQD will not reveal the difference between SQD and DQD. Actually, we take the asymmetric configuration given by $N_1=N$ and $N_2=N_3=1$. Accordingly, we obtain from Eq.(\ref{VarLambda3})
\begin{equation}
\mathrm{rms}[\lambda_{\alpha}]= \mathrm{rms}\left[\frac{I^0_i}{I_1^\alpha}\right] =\sqrt{\frac{1}{16} \times \frac{1}{N^3}}= 0.25 \times \sqrt{\frac{1}{N}} \times \frac{1}{N} ,\label{rmslI}
\end{equation}
whereas, from Eq. (\ref{VarLambda4}),
\begin{equation}
\mathrm{rms}[\lambda_{\alpha}]= \mathrm{rms}\left[\frac{I^0_i}{I_1^\alpha}\right] =\sqrt{\frac{3}{16} \times \frac{1}{N^3}}\approx 0.43 \times \sqrt{\frac{1}{N}} \times \frac{1}{N} .\label{rmslII}
\end{equation}
The Ref.~\cite{Mendes} uses a device with lead width ($W$) in the order of  $\mathrm{mm}$. Therefore, we can estimate that $\sqrt{1/N}\approx \sqrt{\lambda_F/W} \approx \sqrt{\mathrm{nm}/(10^6 \; \mathrm{nm})}\approx \sqrt{10^{-6}}=10^{-3}$, which allow us to rewrite the Eqs.(\ref{rmslI}) and (\ref{rmslII}) as $\mathrm{rms}[\lambda_{\alpha}]\approx 10^{-3} \times 1/N$. Finally, using the same previously dimensional analises to $\lambda_{\mathrm{IERR}}=j^c/j^s$, we obtain $\mathrm{rms}[\lambda_{\mathrm{IERR}}]\approx 10^{-3} \;\mathrm{nm}$, which is in agreement with the Ref.~\cite{Mendes}.  Therefore, we conclude that the difference between experimental measure of Refs. ~\cite{Rojas-Sanchez} and ~\cite{Mendes} is a device effect generated by an geometrically asymmetric configuration leads.

\begin{figure}
\centering
\includegraphics[width=7cm]{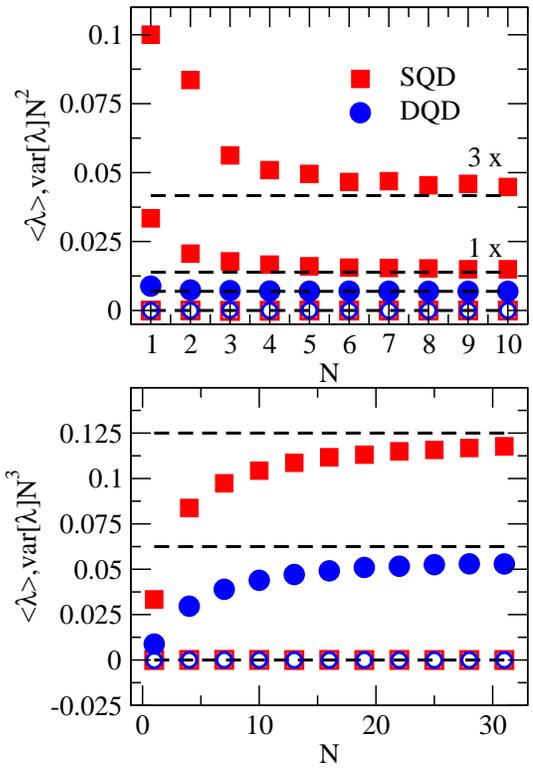}
\caption{The numerical simulation of the SQD and the DQD. The open symbols are average of characteristic coefficient ($\lambda_z$) whereas close symbols are its variance for (up) full symmetric open channels $N_i=N$ and (down) asymmetrical open channels $N_1=N$ and $N_2=N_3=1$. The lines are the Eqs.(\ref{media}), (\ref{VarLambda}), (\ref{VarLambda2}), (\ref{VarLambda3}) and (\ref{VarLambda4}).}
\label{Imagem2}
\end{figure}

\section{ Numerical Simulation} 

In order to confirm the analytical results, Eqs.(\ref{media}), (\ref{VarLambda}), (\ref{VarLambda2}), (\ref{VarLambda3}) and (\ref{VarLambda4}), we perform a numerical simulation founded on the Mahaux-Weidenm\"uller formulation ~\cite{Weidenmuller}. The Scattering Matrix of Eq.(\ref{Smatrix}) is written as a function of both the electronic energy ($\epsilon$) and the Hamiltonian ($\mathcal{H}$) which describe the resonances states inside the ballistic chaotic quantum dot. The general expression is
\begin{eqnarray}
\mathcal{S}=\textbf{1}-2i \pi \mathcal{W}^{\dagger}(\epsilon-\mathcal{H} +i\pi\mathcal{W}\mathcal{W}^{\dagger})^{-1}\mathcal{W}.\label{MW}
\end{eqnarray}
The coupling of the resonances states with the propagating modes  (channels) in the three leads is given by the deterministic matrix $\mathcal{W}=\textbf{(}\mathcal{W}_{1},\mathcal{W}_{2},\mathcal{W}_{3}\textbf{)} $. Moreover, this deterministic matrix satisfies non-direct process, i.e., the orthogonality condition $ \mathcal{W}_{i}^{\dagger}\mathcal{W}_{j}=\frac{1}{\pi}\delta_{i,j}$ holds. 

In the framework of RMT, the Schr\"odinger Hamiltonian, which describe a SQD with reversal-time symmetry and also a large SOI, is a member of the gaussian symplectic ensemble (GSE) ~\cite{Mehta}. Furthermore, its entries has the Gaussian distribution given by 
\begin{eqnarray}
\mathcal{P}( \mathcal{H} )\propto \exp\left\lbrace -\frac{M}{\lambda^{2}}Tr( \mathcal{H} ^{\dagger} \mathcal{H} )\right\rbrace,\nonumber
\end{eqnarray}
where $\lambda=M\Delta/\pi$ is the variance related to the electronic single-particle level spacing, $\Delta$, whereas $M$ is the dimension of the $\mathcal{H}$-matrix and number of resonances states inside of SQD.  To ensure the chaotic regime and consequently the universality of the observables, the number of resonances inside quantum dot is taken as large ($M\gg N_T$)~\cite{RamosBarbosa2}. Using the Eqs.(\ref{tauij}), (\ref{C1}), (\ref{C2}) and (\ref{MW}), we execute the numerical simulations whose results appear in the Fig.(\ref{Imagem2})-up for full symmetric open channels ($N_i=N$) and in the Fig.(\ref{Imagem2})-down for non-symmetrical open channels ($N_1=N$ and $N_2=N_3=1$). Each symbol of the Fig.(\ref{Imagem2}) was obtained through $25\times 10^3$ realizations and with $M=300$. The open square symbols are the average of $\lambda_\alpha$ whereas the close ones are the variance. Moreover, the lines represent the analytical results, Eqs.(\ref{media}), (\ref{VarLambda}) and (\ref{VarLambda2}). For large $N$ the numerical simulations is in great agreement with analytical results.

The massless Dirac Hamiltonian satisfy the following anti-commutation relation ~\cite{Verbaarschot}
$$
\mathcal{H}= -\lambda_{z}\mathcal{H}\lambda_{z}, \quad
\lambda_{z}=
 \left[
\begin{array}{cc}
\textbf{1}_{2M}  & 0\\
0 & -\textbf{1}_{2M}
 \end{array}
 \right],\label{H}
$$
where we interpret the $2M$ of 1's and $-1$'s as the number of atoms in the chaotic graphene quantum dot sub-lattices. The anti-commutation relation above, implies that the Dirac Hamiltonian is 
\begin{eqnarray}
\mathcal{H}=\left(
\begin{array}{cc}
\textbf{0} & \mathcal{T} \\
\mathcal{T}^{\dagger} & \textbf{0}
\end{array}
\right).
\label{H1}
\end{eqnarray}
In the framework of RMT, the massless Dirac Hamiltonian, which describe the DQD with reversal-time symmetry and large SOI,  is a element of the chiral gaussian symplectic ensemble (chGSE) ~\cite{Aguinaldo}. Furthermore, the entries of $\mathcal{T}$-matrix have Gaussian distribution given by
$$
P( \mathcal{T} )\propto \exp\left\lbrace -\frac{2 M}{\lambda^{2}}Tr( \mathcal{T} ^{\dagger} \mathcal{T} )\right\rbrace,
$$
where $\lambda=2M\Delta/\pi$. Using the Eqs.(\ref{tauij}), (\ref{C1}), (\ref{C2}), (\ref{MW}) and (\ref{H1}), we develop  the numerical simulations for DQD that appear in the Fig.(\ref{Imagem2})-up for full symmetric open channels ($N_i=N$) and  the Fig.(\ref{Imagem2})-down for non-symmetrical open channels ($N_1=N$ and $N_2=N_3=1$), which was obtained through $25\times 10^3$ realizations and with $M=300$. The open circle symbols are the average of $\lambda_\alpha$ whereas the close ones are the variance.  Moreover, the lines represent the analytical results, Eqs.(\ref{media}), (\ref{VarLambda3}) and (\ref{VarLambda4}). For large $N$ the numerical simulations are in great agreement with the analytical results, as expected.

\section{Conclusions} 

 We present a complete analytical study of the spin-to-charge conversion in the universal regime for a myriad of mesoscopic devices. We obtain two expressions to the characteristic conversion coefficient, Eqs.(\ref{C1}) and (\ref{C2}), using the  Landauer-B\"uttiker framework that can be applied to any kind of electronic devices. The results are applied to different universal chaotic quantum dots with strong spin-orbit interaction. We use a diagrammatic calculation and obtain analytical expression for characteristic coefficient at Wigner-Dyson and Chiral universal classes. The analytical results are confirmed by numerical simulations in the random matrix theory framework. We connected our analytical results with recent experimental measures of Refs.~\cite{Rojas-Sanchez,Zhang,Mendes} giving an explain about the apparent discrepancies between them. Finally, we show that the difference between experimental measures from 2DGE of Ref. ~\cite{Rojas-Sanchez} and SLG of Ref.~\cite{Mendes} is a geometric device effect generated by asymmetric configuration leads.

\section*{Acknowledgments}
This work was partially supported by CNPq, CAPES and FACEPE (Brazilian
Agencies).

\section*{Appendix. Average and Variance of Characteristic Coefficient}

We show how to obtain the average and variance of the characteristic coefficient $\lambda_\alpha$, Eqs.(\ref{media}) and (\ref{VarLambda}). We take the average of characteristic coefficient, Eq.(\ref{C1}), in the limit of large number of open channels $N_T$
\begin{equation}
\left\langle\lambda_{\alpha}\right\rangle = \left\langle\frac{I_{i}^0}{I_1^\alpha}\right\rangle\approx \frac{\left\langle I_{i}^0\right\rangle}{\left\langle I_1^\alpha\right\rangle} = \frac{\left\langle\tau_{i1}^{00}\tau_{11}^{0\alpha}+(2N_1-\tau_{11}^{00})
\tau_{i1}^{0\alpha}\right\rangle}{\left\langle\tau_{11}^{\alpha0}
\tau_{11}^{0\alpha}-(2N_1-
\tau_{11}^{00})(2N_1-
\tau_{11}^{\alpha\alpha})\right\rangle},
\end{equation}
where $ i=2,3$. The average of $\lambda_\alpha$ can be calculated analytically in the framework of random matrix theory expanding it as a function of $N_T$ as in the following
\begin{equation}
\left\langle\lambda_{\alpha}\right\rangle \approx \frac{\langle{T^c}\rangle}
{\langle{T^s}\rangle}+
\frac{\langle{{\delta}T^c}\rangle{\langle T^s\rangle}-
\langle{{\delta}T^s}\rangle{\langle T^c}\rangle}
{\langle{T^s}\rangle^2}+\mathcal{O}(N_T^{-1})\label{exp}
\end{equation}
where $T^c=\tau_{i1}^{00}\tau_{11}^{0\alpha}+(2N_1-\tau_{11}^{00})\tau_{i1}^{0\alpha}$ and $T^s=\tau_{11}^{\alpha0}
\tau_{11}^{0\alpha}-(2N_1-
\tau_{11}^{00})(2N_1-
\tau_{11}^{\alpha\alpha})$. From the Eq. (\ref{exp}), we can write
\begin{eqnarray}
\langle{T^c}\rangle&=&\langle{\tau_{i1}^{00}
\tau_{11}^{0\alpha}}\rangle+2N_1\langle{\tau_{i1}^{0\alpha}}\rangle-\langle{\tau_{i1}^{0\alpha}\tau_{11}^{00}}\rangle,\nonumber\\
\langle{T^s}\rangle&=&\langle{\tau_{11}^{\alpha0}
\tau_{11}^{0\alpha}}\rangle-4N^2_1+2N_1\langle{\tau_{11}^{\alpha\alpha}}\rangle+2N_1\langle{\tau_{11}^{00}}\rangle-\langle{\tau_{11}^{\alpha\alpha}\tau_{11}^{00}}\rangle,\nonumber
\end{eqnarray}
and
\begin{eqnarray}
\langle{{\delta}T^c}\rangle&=&2\langle{\tau_{j1}^{00}}\rangle\langle{\tau_{11}^{0\alpha}}\rangle+2N_1\langle{\tau_{j1}^{0\alpha}}\rangle-2\langle{\tau_{j1}^{0\alpha}}\rangle\langle{\tau_{11}^{00}}\rangle,\nonumber\\
\langle{{\delta}T^s}\rangle&=&2\langle{\tau_{11}^{\alpha0}}\rangle\langle{\tau_{11}^{0\alpha}}\rangle+2N_1\langle{\tau_{11}^{00}}\rangle+2N_1\langle{\tau_{11}^{\alpha\alpha}}\rangle-2\langle{\tau_{11}^{00}}\rangle\langle{\tau_{11}^{\alpha\alpha}}\rangle.\nonumber
\end{eqnarray}
For a chaotic Schr\"{o}dinger quantum dot in the presence of  time-reversal symmetry and strong spin-orbit interaction, we can use the standard diagrammatic method of Refs. ~\cite{Jacquod,Brouwer,RamosBarbosa2}. Performing the calculations, we obtain
\begin{eqnarray}
\langle{T^c}\rangle&=&0,\label{A1}\\
\langle{T^s}\rangle&=&-\frac{4N_1^2(N_2+N_3)}{N_T},\label{B1}\\
\langle{{\delta}T^c}\rangle&=&0,\label{A2}\\
\langle{{\delta}T^s}\rangle&=&\frac{8N_1^2(2N_1^2+2N_1N_2+2N_1N_3-N_1-2N_2-2N_3)}{(2N_T-1)^2}.\label{B2}
\end{eqnarray}

We replace the Eqs.(\ref{A1}-\ref{B2}) in Eq.(\ref{exp}) and obtain
\begin{eqnarray}
\langle{\lambda_{\alpha}}\rangle=0.\nonumber
\end{eqnarray}
Furthermore, we can do the same procedure to obtain the variance of $\lambda_\alpha$. The variance is defined as
\begin{eqnarray}
\mathrm{var}[\lambda_{\alpha}]=\langle\lambda_{\alpha}^2\rangle-\langle\lambda_{\alpha}\rangle^2=\langle\lambda_{\alpha}^2\rangle .\nonumber
\end{eqnarray}
Expanding in order of $\mathcal{O}(N_T^{-1})$, we have that
\begin{eqnarray}
\mathrm{var}[\lambda_{\alpha}]&=&\langle\lambda^2_{\alpha}\rangle \approx \frac{\langle{T^c}\rangle^2}
{\langle{T^s}\rangle^2}\nonumber\\&+&
\frac{\langle{{\delta}{T^c}^2}\rangle\langle{T^s}\rangle^2 +\langle{{\delta}{T^s}^2}\rangle{\langle{T^c}\rangle}^2-2\langle{{\delta}T^c{\delta}T^s}\rangle{\langle{T^c}\rangle}\langle{T^s}\rangle}
{\langle{T^s}\rangle^4}\nonumber\\&+&2\frac{\langle{{\delta}T^c}\rangle\langle{T^c}\rangle{\langle{T^s}\rangle}-
\langle{{\delta}T^s}\rangle\langle{T^c}\rangle^2}
{\langle{T^s}\rangle^3}+\mathcal{O}(N_T^{-1}).\label{var}
\end{eqnarray}
From Eqs. (\ref{A1}-\ref{B2}), the Eq. (\ref{var}) simplify to 
\begin{equation}
\mathrm{var}[\lambda_{\alpha}]\approx\frac{\langle{{\delta}{T^c}^2}\rangle}{\langle{T^s}\rangle^2}.\label{var2}
\end{equation} 
Using Refs. ~\cite{Jacquod,Brouwer,RamosBarbosa2}, we obtain that
\begin{equation}
\langle{{\delta}{T^c}^2}\rangle=\frac{4N_1^3N_2N_3(N_2+N_3)}{N_T^4}\label{dA2}
\end{equation}
Replacing the Eqs. (\ref{B1}) and (\ref{dA2}) in Eq. (\ref{var2}), finally we obtain that
\begin{equation}
\mathrm{var}[\lambda_{\alpha}]=\frac{N_2N_3}{4N_1(N_2+N_3)N^2_T}.
\end{equation}
The Eqs. (\ref{VarLambda2}), (\ref{VarLambda3}) and (\ref{VarLambda4}) can be obtained following the same procedure.

\end{document}